OPEN ACCESS

International Journal of Poultry Science

ISSN 1682-8356
DOI: 10.3923/ijps.2020.346.355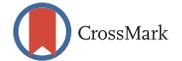

## Research Article
## Phenotypic Correlation Between External and Internal Egg Quality Characteristics in 85-Week-Old Laying Hens

[1]J.S. Inca, [2]D.A. Martinez and [1]C. Vilchez

[1]Department of Nutrition, Universidad Nacional Agraria La Molina, Lima, Peru
[2]Center of Excellence for Poultry Science, University of Arkansas, Fayetteville, AR 72701, USA## Abstract

**Background and Objective:** Several studies confirm that the age of hens has a tremendous impact on external and internal egg quality characteristics. Egg production could be at serious risk if egg quality characteristics and age of hens are not seriously considered. This study was conducted to analyze the phenotypic correlations between some internal and external egg quality characteristics in old laying hens. **Materials and Methods:** A total of 288 eggs of 85-week-old Hy-Line Brown laying hens were collected during 3 weeks and their internal and external egg characteristics were evaluated. **Results:** Phenotypic correlations between egg quality characteristics in old laying hens indicate a negative impact on shell and albumen quality but not affected yolk quality characteristics. **Conclusion:** This study helps to understand that raising laying hens above 80 weeks would have a negative impact on egg quality characteristics.

Citation: J.S. Inca, D.A. Martinez and C. Vilchez, 2020. Phenotypic correlation between external and internal egg quality characteristics in 85-week-old laying hens. Int. J. Poult. Sci., 19: 346-355.

Corresponding Author: C. Vilchez, Department of Nutrition, Universidad Nacional Agraria La Molina, Lima, PeruCopyright: © 2020 J.S. Inca *et al.* This is an open access article distributed under the terms of the creative commons attribution License, which permits unrestricted use, distribution and reproduction in any medium, provided the original author and source are credited.Competing Interest: The authors have declared that no competing interest exists.

Data Availability: All relevant data are within the paper and its supporting information files.

**INTRODUCTION**

Worldwide, more than 7 billion hens are laying around 1.3 trillion eggs per year[1]. In addition, from 1960 to 2010, the worldwide egg production has shown an increase of 30%[2]. Thus, egg quality have gained great relevance in the egg production process. Many egg quality characteristics directly affect consumer's acceptability[3]. Egg quality comprises a number of aspects related to the shell, albumen and yolk and can be divided into external and internal quality[4]. Both the reproductive process and consumption have a strong connection with egg quality characteristics. Altinel et al.[5] reported that external and internal egg quality characteristics are relevant in poultry breeding, because of their effects on reproductive output and growth of progeny. Regarding egg conservation, It has been reported that 8% of egg production break during transport from farms to consumers[6], these and the number of eggs cracked are a serious economic issue for both breeders and distributors[7]. Understanding correctly egg quality characteristics may diminish the rate of loss during commercial processing, considerably[8]. Previous studies have shown relationship between external and internal egg quality characteristics and these relations are affected by age, genotype of hen, type of rearing system and nutrition[9,10]. Therefore, it is important to pay attention to these characteristics in order to maintain the quality and avoid problems of preservation and marketing of eggs[3].

Past investigations stated that the age of the bird has a direct effect on egg quality characteristics[11] and may change these characteristics in size, weight and ratio[9]. External egg characteristics are strongly affected by age of hens. Average egg weight increases with the increase of the breeder age[12,13]. Silverside and Scott[14] reported a negative correlation between age of hen and shell quality characteristics. Shell quality decreases with advancing age, which is presumably related to the increasing of egg size and egg surface area[15]. Other studies informed that young hens produce eggs with shell thicker than old hens[8]. Increasing the age of hen not only affects the external egg characteristics but also has an impact on internal quality, which decreases[14,16]. Egg weight elevation increases both yolk and albumen weight. However, there is a greater proportion of yolk and a smaller proportion of albumen noted with the increase of the breeder age[13,17]. Lapão et al.[18] indicated that albumen height was affected by age of hen, from 8.80-8.30 for 32 and 59 weeks breeder age, respectively. In addition, Haugh unit presented a reduction with the increase of age of hen, from 88.6-82.1 at hens for 35 and 45 weeks of age, respectively[12]. Furthermore, Hy-Line[19] has indicated that hen-egg rate of Hy-Line Brown laying hens diminishes until 75% or less since 80 weeks of age. The study reported herein was conducted to determine the phenotypic correlation between external and internal egg quality characteristics in 85-week-old laying hens, establishing the implications of these correlations on egg quality and production at this hen stage.

**MATERIALS AND METHODS**

**Biological materials:** A total of 288 eggs from 14485-week-old Hy-Line Brown laying hens belonging to the Poultry Experimental Unit of the Animal Science Department, Universidad Nacional Agraria La Molina were used. All hens were raised in the same conditions (temperature, humidity, feed distribution, feed time, water disposition, etc.). The eggs used for the present study were sampled within a 3-week period, one day at each week samples was taken and all egg laid during 24 hours of the day were evaluated. (90, 97 and 101 eggs at 85, 86 and 87 weeks of age, respectively). The eggs collected were evaluated using various egg quality tests for both external and internal characteristics.

**Egg quality determination:** All collected eggs were labeled with consecutive numbers in order to trace every egg during the whole quality evaluation process. Egg quality evaluation process started with the evaluation of egg weight (EW), for this purpose an electronic centesimal scale with a 300 g capacity was used. Then, a digital caliper was used to measure egg length (EL) and width (EG). Finally, a densimeter, 10 buckets of 20 L capacity and salt (saline solution technique) were used to determine specific gravity as described by Bennett[20]. Subsequently, the eggs were broken on a smooth plastic platform (previously calibrated) and the albumen and yolk weights, lengths and heights (albumen weight (AW), albumen length (AL), albumen height (AH), yolk weight (YW), yolk length (YL) and yolk height (YH), respectively) were determined using a digital caliper. Afterward, the shells were washed and stored in a plastic container (10×10 cm) for drying at room temperature (21.5°C). Finally, after 3 days, shell weight (SW) and shell thickness (ST) were measured.

Measured egg quality characteristics data were used to calculate some external and internal egg quality characteristics. These calculated characteristics were estimated using equations obtained from Kul and Seker[6], Romanoff and Romanoff[21], Paganelli et al.[22], Singh[23], Alkan et al.[24] and Debnath and Ghosh[25].

**External egg characteristics:**

$$\text{Egg shape index (ESI, \%)} = \frac{\text{EG}}{\text{EL}} \times 100$$

where, EG and EL are the egg width and length, respectively.

$$\text{Egg surface area (ESA, cm}^2\text{)} = 3.9782 \times EW^{0.75056}$$

where, EW is the egg weight (g).

$$\text{Unit surface shell weight}\left(\frac{U, mg}{cm^2}\right) = \frac{SW}{ESA}$$

Where
SW = Shell weight (mg)
ESA = Egg surface area (cm$^2$)

$$\text{Shell ratio (SR, \%)} = \frac{SW}{EW} \times 100$$

**Internal egg characteristics:**

$$\text{Albumen index (AI, \%)} = \frac{AH}{[DAL + DAG]/2} \times 100$$

Where
AH = Albumen height (mm)
DAL = Dense albumen length (mm)
DAG = Dense albumen width (mm)

$$\text{Albumen ratio (AR, \%)} = \frac{AW}{EW} \times 100$$

where, AW and EW are the albumen (g) and egg weights (g), respectively.

$$\text{Haugh unit (HU)} = 100 \times \log(AH - 1.7EW^{0.37} + 7.6)$$

Where
AH = Albumen height (mm)
EW = Egg weight (g)

$$\text{Yolk index (YI, \%)} = \frac{YH}{[YL + YG]/2} \times 100$$

Where
YH = Yolk height (mm)
YL = Yolk length (mm)
YG = Yolk width (mm)

$$\text{Yolk ratio (YR \%)} = \frac{YW}{EW} \times 100$$

where, YW and EW are the yolk and egg weight, respectively.

$$\text{Yolk - albumen ratio (Y : A, \%)} = \frac{YW}{AW} \times 100$$

where, YW and AW are the yolk and albumen weights (g), respectively.

**Statistical analysis:** Data were analyzed using a one-way ANOVA to obtain residuals. Anderson-Darling test was used to analyze the normality of the residuals of all variables evaluated using Minitab 16 Statistical Program[26]. Correlation analysis of egg characteristics was obtained with Pearson product-moment correlation coefficients (PCC) using SPSS 20.0 computerpackageprogram[27]. $p<0.05$ and $p<0.01$ were considered significant and highly significant, respectively. The coefficients obtained from the Pearson correlation analysis were interpreted as indicated in Table 1.

Table 1: Interpretation table of the Pearson correlation coefficient[1]

| Correlation coefficient[2] | Interpretation of r |
|---|---|
| -1.00 | Perfect negative correlation: ("A major X, minor Y", proportionally. It means, every time X increases a unit, Y always decreases a constant amount). This also applies "A minor X, greater Y"[3] |
| -0.90 | Very strong negative correlation |
| -0.75 | Considerable negative correlation |
| -0.50 | Medium negative correlation |
| -0.25 | Weak negative correlation |
| -0.10 | Very weak negative correlation |
| 0.00 | There is no correlation between the variables |
| +0.10 | Very weak positive correlation |
| +0.25 | Weak positive correlation |
| +0.50 | Medium positive correlation |
| +0.75 | Considerable positive correlation |
| +0.90 | Very strong positive correlation |
| +1.00 | Perfect positive correlation: ("A major X, greater Y" or "a minor X, minor Y", proportionally, every time X increases, Y always increases a constant amount) |

[1]Source: Hernández *et al.*[28] [2]- or +: Direction of the correlation, 1.00: Magnitude of the correlation. [3]X: Independent variable, Y: Dependent variable

## RESULTS

**Descriptive statistic:** Table 2 presents the descriptive statistics obtained from the data on external and internal egg characteristics. The average EW, EL, EG, SW, ST, SR, ESI, ESA, U and Specific gravity were 69.79 g, 61.36 mm, 44.81 mm, 6.28 g, 0.35 mm, 9.01%, 73.14%, 96.24 cm$^2$, 65.24 mg/cm$^2$ and 1.088, respectively. In addition, the average AW, AL, AG, DAL, DAG, AH, AR, AI and HU were 43.19 g, 124.7 mm, 101.9 mm, 92.46 mm, 80.75 mm, 7.10 mm, 61.81%, 8.39% and 80.48, respectively. Furthermore, the average YW, YL, YG, YH, YR, YI and Y:A were 17.17 g, 43.02 mm, 40.93 mm, 15.46 mm, 24.64%, 36.88% and 40.08%, respectively.

**Phenotypic correlations among external egg characteristics:** Table 3 shows the phenotypic correlations between external egg characteristics. A highly significant, weak and negative (p<0.01; -0.15) phenotypic correlation was found between EW and ESI. Likewise, a significant, weak and negative phenotypic correlation (p<0.05; -0.15) was found

Table 2: Descriptive statistics of egg quality characteristics of 85-week-old laying hens[1]

| Characteristics | Mean | SEM[2] | Minimum | Maximum | CV[3] |
|---|---|---|---|---|---|
| **External egg quality characteristics** | | | | | |
| Egg weight (g) | 69.79 | 0.319 | 58.520 | 87.310 | 7.75 |
| Egg length (mm) | 61.36 | 0.157 | 55.700 | 71.010 | 4.35 |
| Egg width (mm) | 44.81 | 0.075 | 41.440 | 48.160 | 2.83 |
| Egg shape index (%) | 73.14 | 0.190 | 62.210 | 80.260 | 4.41 |
| Shell weight (g) | 6.280 | 0.043 | 3.220 | 9.000 | 11.54 |
| Shell thickness (mm) | 0.349 | 0.003 | 0.230 | 0.490 | 12.21 |
| Shell ratio (%) | 9.008 | 0.052 | 5.350 | 11.900 | 9.85 |
| ESA[4] (cm$^2$) | 96.24 | 0.329 | 84.360 | 113.910 | 5.81 |
| U[5] (mg/cm$^2$) | 65.24 | 0.375 | 37.380 | 87.970 | 9.75 |
| Specific gravity | 1.088 | 0.000 | 1.055 | 1.110 | 0.63 |
| **Internal egg quality characteristics** | | | | | |
| Albumen weight (g) | 43.19 | 0.264 | 33.720 | 61.020 | 10.37 |
| Total albumen length (mm) | 124.7 | 0.936 | 81.690 | 193.940 | 12.74 |
| Total albumen width (mm) | 101.9 | 0.823 | 43.160 | 154.340 | 13.71 |
| Dense albumen length (mm) | 92.46 | 0.625 | 70.500 | 145.100 | 11.48 |
| Dense albumen width (mm) | 80.75 | 0.564 | 41.960 | 125.560 | 11.85 |
| Albumen height (mm) | 7.103 | 0.075 | 2.930 | 11.120 | 17.90 |
| Albumen ratio (%) | 61.81 | 0.164 | 53.980 | 72.510 | 4.51 |
| Albumen index (%) | 8.396 | 0.125 | 2.730 | 15.760 | 25.25 |
| Haugh unit | 80.48 | 0.574 | 39.100 | 102.710 | 12.10 |
| Yolk weight (g) | 17.17 | 0.109 | 8.080 | 25.820 | 10.74 |
| Yolk length (mm) | 43.02 | 0.107 | 37.720 | 48.960 | 4.22 |
| Yolk width (mm) | 40.93 | 0.130 | 27.860 | 45.430 | 5.41 |
| Yolk height (mm) | 15.46 | 0.046 | 13.450 | 17.240 | 5.01 |
| Yolk ratio (%) | 24.64 | 0.133 | 13.730 | 33.780 | 9.13 |
| Yolk index (%) | 36.88 | 0.121 | 30.360 | 46.840 | 5.57 |
| Yolk: Albumen ratio (%) | 40.08 | 0.307 | 18.930 | 62.580 | 13.00 |

[1]Total eggs evaluated: n = 288. [2]SEM: Standard error mean (%). [3]CV: Coefficient of variation (%). [4]ESA: Egg surface area. [5]U: Unit surface shell weight

Table 3: Phenotypic correlations of external egg quality characteristics of 85-week-old laying hens[1]

| External characteristics | Egg length | Egg width | Shell weight | Shell thickness | Shell ratio | ESI[2] | ESA[3] | U[4] | Specific gravity |
|---|---|---|---|---|---|---|---|---|---|
| Egg weight | 0.72** | 0.84** | 0.54** | 0.11 | -0.15* | -0.15** | 1.00** | 0.05 | -0.07 |
| Egg length | | 0.27** | 0.26** | -0.01 | -0.26** | -0.79** | 0.72** | -0.12* | -0.25** |
| Egg width | | | 0.46** | 0.07 | -0.11 | 0.38** | 0.84** | 0.05 | -0.02 |
| Shell weight | | | | 0.70** | 0.75** | 0.04 | 0.54** | 0.87** | 0.70** |
| Shell thickness | | | | | 0.73** | 0.05 | 0.11 | 0.76** | 0.69** |
| Shell ratio | | | | | | 0.17** | -0.15* | 0.98** | 0.88** |
| ESI | | | | | | | -0.15** | 0.14* | 0.22** |
| ESA | | | | | | | | 0.05 | -0.07 |
| U | | | | | | | | | 0.87** |

[1]Total eggs evaluated: n = 288. [2]ESI: Egg shape index. [3]ESA: Egg surface area. [4]U: Unit surface shell weight. *p<0.05

between EW and SR. In addition, EW and specific gravity showed a negative phenotypic correlation of -0.07. The phenotypic correlation between EW and SW was medium, positive (0.54) and highly significant (p<0.01), while that between EW and ST was not significant (p>0.05; 0.11).

We found weak, negative phenotypic correlations of EL with SR (-0.26) and specific gravity (-0.25); both correlations being highly significant (p<0.01).Correlations of EL and EG with ESI showed different results: EL and ESI showed a strong negative phenotypic correlation of -0.79, while the EG and ESI showed a medium positive phenotypic correlation of 0.38.

The phenotypic correlation between SR and ESI was weak and positive (0.17), while that between SR and ESA was weak and negative (-0.15); both correlations being highly significant (p<0.01). In addition, specific gravity showed a medium positive phenotypic correlation with SW (0.70) and ST (0.69). In addition, it showed a considerable positive phenotypic correlation with SR (0.88) and U(0.87). Both correlations were highly significant (p<0.01).

**Phenotypic correlations among internal egg characteristics:**
Table 4 shows the phenotypic correlations between internal egg characteristics. Highly significant, weak and positive phenotypic correlations were found between AW and albumen dimensions (TAL, TAW, DAL and DAW) (p<0.01). Also, highly significant, weak and negative phenotypic correlations were obtained of AW with AH (-0.15), AI (-0.27) and HU (-0.29) (p<0.01). The phenotypic correlations of AW with YW, YL, YG, YH and YI were highly significant, weak and positive. Furthermore, highly significant (p<0.01), medium and negative phenotypic correlations were found between AW with the YR (-0.52) and Y:A (-0.60), respectively.

The highly significant (p<0.01) medium and negative phenotypic correlation coefficients were found between the albumen dimensions (TAL, TAW, DAL and DAW) and AH, AI and HU. In this study, no significant correlations were found between albumen dimensions and yolk characteristics. In contrast, the AR showed weak and negative correlations with AH, AI, HU, YL, YG and YH, while the correlation between AR with YR and Y:A was negative and strong (-0.85 and -0.93, respectively). All these correlations were significant (p<0.01). Moreover, highly significant, medium and positive phenotypic correlations were obtained between YW with YL (0.62), YG (0.59), YH (0.50), YR (0.71) and Y:A (0.65) (p<0.01).

**Phenotypic correlations between external and internal egg characteristics:** Table 5 shows the phenotypic correlations between external and internal egg characteristics. Highly

Table 4: Phenotypic correlations of internal egg quality characteristics of 85-week-old laying hens.[1]

| Internal characteristics | TAL[2] | TAW[3] | DAL[4] | DAW[5] | Albumen height | Albumen ratio | Albumen index | Haugh unit | Yolk weight | Yolk length | Yolk width | Yolk height | Yolk ratio | Yolk index | Y/A[6] |
|---|---|---|---|---|---|---|---|---|---|---|---|---|---|---|---|
| Albumen weight | 0.41** | 0.46** | 0.40** | 0.41** | -0.15* | 0.71** | -0.27** | -0.29** | 0.20** | 0.16** | 0.16** | 0.28** | -0.52** | 0.11 | -0.60** |
| TAL | | 0.60** | 0.48** | 0.37** | -0.55** | 0.31** | -0.56** | -0.59** | 0.08 | 0.01 | 0.01 | 0.07 | -0.21** | 0.05 | -0.25** |
| TAW | | | 0.55** | 0.57** | -0.64** | 0.31** | -0.68** | -0.70** | 0.12* | 0.04 | 0.03 | 0.10 | -0.22** | 0.06 | -0.25** |
| DAL | | | | 0.77** | -0.69** | 0.27** | -0.82** | -0.72** | 0.06 | 0.11 | 0.11 | 0.07 | -0.24** | -0.03 | -0.25** |
| DAW | | | | | -0.61** | 0.18** | -0.76** | -0.66** | 0.22** | 0.20** | 0.18** | 0.13* | -0.11 | -0.05 | -0.14* |
| Albumen height | | | | | | -0.10 | 0.96** | 0.97** | 0.01 | -0.04 | -0.03 | -0.01 | 0.11 | 0.02 | 0.10 |
| Albumen ratio | | | | | | | -0.17** | -0.16** | -0.46** | -0.24** | -0.24** | -0.12* | -0.85** | 0.09 | -0.93** |
| Albumen index | | | | | | | | 0.94** | -0.06 | -0.11 | -0.09 | -0.05 | 0.15* | 0.04 | 0.16** |
| Haugh unit | | | | | | | | | -0.08 | -0.07 | -0.07 | -0.07 | 0.15* | -0.01 | 0.15* |
| Yolk weight | | | | | | | | | | 0.62** | 0.59** | 0.50** | 0.71** | -0.08 | 0.65** |
| Yolk length | | | | | | | | | | | 0.60** | 0.27** | 0.41** | -0.44** | 0.36** |
| Yolk width | | | | | | | | | | | | 0.30** | 0.41** | -0.47** | 0.19** |
| Yolk height | | | | | | | | | | | | | 0.23** | 0.65** | 0.98** |
| Yolk ratio | | | | | | | | | | | | | | -0.15* | -0.14* |

[1]Total eggs evaluated: n = 288. [2]TAL: Total albumen length. [3]TAW: Total albumen width. [4]DAL:Dense albumen length. [5]DAW:Dense albumen width. [6]Y/A: Yolk-albumen ratio. * p<0.05, **p<0.01

Table 5: Phenotypic correlations between external and internal egg characteristics of 85-week-old laying hens[1]

| Characteristics | Egg weight | Egg length | Egg width | Shell weight | Shell thickness | Shell ratio | ESI[2] | Specific gravity |
|---|---|---|---|---|---|---|---|---|
| Albumen weight | 0.91** | 0.70** | 0.75** | 0.39** | -0.01 | -0.25** | -0.19** | -0.16** |
| TAL[3] | 0.37** | 0.43** | 0.20** | 0.20** | 0.00 | -0.05 | -0.28** | -0.03 |
| TAW[4] | 0.43** | 0.42** | 0.29** | 0.22** | 0.07 | -0.07 | -0.22** | -0.05 |
| DAL[5] | 0.37** | 0.43** | 0.20** | 0.28** | 0.06 | 0.05 | -0.28** | 0.04 |
| DAW[6] | 0.43** | 0.40** | 0.29** | 0.30** | 0.08 | 0.02 | -0.20** | 0.00 |
| Albumen height | -0.13* | -0.30** | 0.03 | -0.19** | -0.15* | -0.12* | 0.30** | -0.10 |
| Albumen ratio | 0.37** | 0.35** | 0.26** | -0.03 | -0.21** | -0.32** | -0.17** | -0.23** |
| Albumen index | -0.26** | -0.39** | -0.07 | -0.25** | -0.13* | -0.10 | 0.32** | -0.08 |
| Yolk weight | 0.54** | 0.32** | 0.51** | 0.35** | 0.14* | -0.01 | 0.00 | 0.01 |
| Yolk length | 0.37** | 0.25** | 0.33** | 0.26** | 0.09 | 0.02 | -0.04 | -0.02 |
| Yolk width | 0.35** | 0.25** | 0.30** | 0.27** | 0.12* | 0.04 | -0.05 | 0.04 |
| Yolk height | 0.43** | 0.20** | 0.44** | 0.37** | 0.15* | 0.09 | 0.08 | 0.09 |
| Yolk ratio | -0.20** | -0.22** | -0.11 | -0.04 | 0.07 | 0.11 | 0.13* | 0.06 |
| Yolk index | 0.08 | -0.03 | 0.12* | 0.10 | 0.04 | 0.05 | 0.12 | 0.06 |

[1]Total eggs evaluated: n = 288. [2]ESI: Egg shape index. [3]TAL: Total albumen length. [4]TAW: Total albumen width. [5]DAL: Dense albumen length. [6]DAW: Dense albumen width. *p<0.05

significant phenotypic correlations (p<0.01) were found between EW with AW (strong positive, 0.91) and YW (medium positive, 0.54). Furthermore, highly significant (p<0.01), weak and positive phenotypic correlations were found between EW and albumen and yolk dimensions. Highly significant (p<0.01) phenotypic correlations were found between EW with AR and YR, this correlations were weak and positive (0.37) and weak and negative (-0.20), respectively.

Phenotypic correlations between EL with AW and YW were highly significant (p<0.01), medium positive (0.70) and weak positive (0.32), respectively. Furthermore, EL was correlated positively and weakly with AR (0.35) and YH (0.20) but negatively and weakly with AH (-0.30), AI (-0.39) and YR (-0.22). Phenotypic correlations among EG with AW and YW were considerable positive (0.75) and medium positive (0.51), respectively. All these correlations were highly significant (p<0.01).

Phenotypic correlations between SW with internal egg characteristics (AW, YW, YL, YG and YH) were weak positive and highly significant (p<0.01). Meanwhile, a highly significant (p<0.01) weak negative phenotypic correlation between SW with AH and AI were found.

## DISCUSSION

**Phenotypic correlations among external egg characteristics:** Studies indicated that EW increases with age of the bird[12,13]. Therefore, the results support the earlier findings that decrease in ESI is poorly affected when EW increase[4]. Also, Kul and Seker[6], Sezer[29] and Bernacki et al.[30] reported similar correlations between EW and ESI than in the current investigation. Mitrovic et al.[31] reported that egg weight of old laying hen had a negative impact on SR but this effect was minimum. Bernacki et al.[30] also reported a significant, weak and negative phenotypic correlation (p<0.05; -0.17) between EW and SR. Many studies have shown similar findings[6,16,30,32]. Phenotypic correlation between EW and specific gravity indicate that there is actually no linear correlation between these characteristics. The increase of EW (because of age of hen) was not proportioned with the increase of SW, as compared to young hens where SW increases proportionally with EW[33]. Earlier studies indicated a reduction in ST when EW increases[34-36]. However, Agaviezor et al.[37] reported the importance of the correlation between EW with ST and SW, because to determine the ST and SW it is necessary to break the egg and spend a lot of time to realize subsequent measurements, which decreases the efficiency of egg production.

Correlations between EL with SR and specific gravity indicate that the larger the egg, the lower the shell quality. Alkan et al.[33] reported that an increase in EL affects shell quality negatively, because egg size does not increase linearly with shell percentage in old laying hens. The present study agrees with the findings of Altuntaş and Şekeroğlu[38] who stated that the advanced age of hens had no effect on egg size measurements, ESI and their correlation. Moreover, phenotypic correlations of ESI with EL and EG found in the present study agree with those of Kul and Seker[6], who found phenotypic correlations of ESI with EL (-0.77) and EW (0.34) in Japanese quails. Thus, ESI is not affected by advanced age of hen, EW and even by the specie of laying bird.

Results indicate that advanced age of hen reduces SR but such reduction has no effect on ESI and ESA. The phenotypic correlation between SR and ESA found in the present study is different from those of Sezer[29] and Fajemilehin et al.[39], because previous studies were conducted with younger birds. Correlations of specific gravity with SW, ST, SR and U showed that specific gravity is a good indicator of shell quality because it has a high and direct connection with principal shell characteristics. But, specific gravity tests may increase the cost of the egg quality determination process.

According to correlations of the external egg quality characteristics, an egg produced by old laying hens (more than 70 weeks in Hy-Line) can diminish shell quality; plus, Hy-Line[19] indicated that hen-egg rate of Hy-Line Brown hens diminishes until 75% or less since 80 weeks of age. Both factors reduce the efficiency of egg production and compromise the profits of egg farmers.

**Phenotypic correlations among internal egg characteristics:** Albumen dimensions measurement results indicated that TAL, TAW, DAL and DAW slightly increase with an increment in AW. No studies have reported phenotypic correlations between these characteristics before. The current outcomes indicate thin albumen increased when age of hen increases; therefore, both TAL and TAW increased more than DAL and DAW. This result suggests that albumen quality might diminish in old laying hens. Jacob et al.[40] mentioned that increase in thin albumen is very affected by age of hen, affecting albumen quality.

Correlations between AW with AH and UH found in this study agrees with previous investigations[41-43]. AW increase with age of hen but AH and HU have a negative impact[44], because, albumen pH increases with increasing age of hen[45]. HU is a good indicator of albumen quality; so, a very low HU in 85-week-old hens showed a poor albumen quality. Moreover, correlations between AW and yolk quality characteristics indicated that an increase in AW slightly increase the YW, YL, YG, YH and YI. No past studies correspond with these correlations outcomes[25,39,46,47]. In addition, Bernacki et al.[30] and Seker[48] reported similar results and found highly significant (p<0.01), medium and positive correlations of AW with the yolk ratio (-0.52) and Y:A (-0.58). In the present study, high EW (because of the age of hen) have a greater proportion of albumen but a lower proportion of yolk, which was in agreement with the results of Hussein et al.[41].

The results of correlations between albumen dimensions (TAL, TAW, DAL and DAW) with AH, AI and HU found in the present study are different from those of Alkan et al.[24], who presented weak negative phenotypic correlations of albumen dimensions with AH and HU, respectively. Proportional improvements in all albumen characteristics were found in young hens (from 30-50 weeks of age) but this improvement is not continuous in old laying hens. Thick albumen increase in length dimensions but no in height[44].

An increase in AR of aged hens has a minimum impact on albumen and yolk characteristics. Previous findings mentioned that internal egg components increase in different proportions due to age of hens[44,49]. Kul and Seker[6] reported a similar correlation between AR and YR (-0.95) in young quails compared to the current study. Correlations between albumen and yolk ratios found in the present study showed an inconsistency with previous findings, Y:A tended to be greater in smaller eggs (young hens) than in larger eggs (old hens)[50].

The phenotypic correlations between YW and yolk characteristics are consistent with the results obtained in past researches. For instance, Kul and Seker[6] reported a medium positive phenotypic correlations(0.55)between YW and yolk diameters (length and width). Ojo et al.[51] reported a medium positive phenotypic correlations (0.64) between YW and YH. In addition, Olawumi and Ogunlade[46] reported a medium positive phenotypic correlation (0.64) between YW and YR. Alkan et al.[24] found a positive phenotypic correlation (0.67) between YW and Y:A; in addition, a highly significant, positive phenotypic correlation (0.98) (p<0.01) between YR and Y:A. This result suggests that there is an almost perfect relationship between the two characteristics. Alkan et al.[24] found a similar highly significant phenotypic correlation between these characteristics (0.97; p<0.01). These similarities were obtained despite the other investigations were carried out on young hens; therefore, the advanced age of hens (80 weeks) not alter yolk quality characteristics and theircorrelations[14,42,50,52].

**Phenotypic correlations between external and internal egg characteristics:** Correlations of EW with AW and YW indicate that EW increases (due to age of hen) and have a high connection with a change in AW but, this would not have the same impact in YW. In addition, Bernacki et al.[30], Agaviezor et al.[37], Emamgholi et al.[53], Alipanah et al.[54] and Shafey et al.[55] reported highly significant (p<0.01) and positive phenotypic correlations between EW and AW and medium positive phenotypic correlations between EW and YW. The advanced age of hens does not alter the YW[50].

Phenotypic correlations of EW with albumen and yolk dimensions showed that EW elevation has a low impact on albumen and yolk dimensions. There are no studies on the correlation between EW and albumen dimensions that agree with the present study. Olawumi and Ogunlade[46], Alkan et al.[24], Debnath et al.[25] and Onunkwo and Okoro[56] indicated similar outcomes between EW and yolk diameter. Again, a high EW (because of the age of hen) do not affect yolk quality characteristics. Similarly, correlations of EW with AR and YR were reported in previous studies[6,30,46,57]. These researchers reported both weak positive and weak negative phenotypic correlations between EW with AR and YR, respectively. An EW elevation has a very low effect on albumen and yolk proportion.

Phenotypic correlations of EL with AW and YW indicate that an increase in egg size affects considerably AW but has a low effect on YW. Olawumi and Ogunlade[46] reported similar findings, a medium positive phenotypic correlation of 0.64 between EL and AW and a weak positive phenotypic correlation of 0.34 between EL and YW. In advanced age of hens, egg size elevation does not alter yolk characteristics. Alike findings were reported by researchers who found correlation between EL and internal egg characteristics. For instance, Alipanah et al.[54] reported a phenotypic correlation of 0.33 and -0.29 between EL with AR and YR. In addition, Albrecht[58] found a weak negative phenotypic correlation (-0.49) between EL and AH. Kul and Seker[6] obtained a weak positive correlation (0.20) between EL and YH. Nonetheless, all these similarities were found in different young laying species. From the results in this study, it may observe that egg size (because of age of hen) has more impact on albumen quality characteristics than yolk quality characteristics.

Olawumi and Ogunlade[46] reported similar correlations between EG with AW and YW, considerable positive (0.77) and medium positive (0.48) phenotypic correlation were found between EG with AW and YW. Current study findings suggest that egg dimension influence in different magnitude both albumen and yolk weight. In young hens, egg dimensions may increase YW; meanwhile, in advanced age hens, YW shows a reduction and AW is not altered in both stages.

Corresponding findings were reported in previous studies. Olawumi and Ogunlade[46] indicated weak positive and highly significant (p<0.01) phenotypic correlations between SW and AW, YW and YG. In addition, Bernacki et al.[30] showed a weak positive phenotypic correlation between SW and YW. SW did not alter internal egg characteristics in old laying hens.

## CONCLUSION

The egg quality characteristics of old laying hens have a negative impact on shell and albumen quality but do not affect on yolk quality characteristics. Old laying hens may diminish shell quality characteristics, such as shell weight, shell thickness and shell ratio, because of the increasing egg weight and size (a well-known egg characteristic in old laying hens). In addition, a clear reduction in albumen quality characteristics was observed in advanced age of hens. Both Haugh unit and albumen height (key egg quality characteristics) decreased their measurements, indicators of a poor albumen quality. Finally, yolk quality characteristics were not affected. This study helps to understand that raising laying hens above 80 weeks would have a negative impact on egg quality characteristics; compromising both the quality and efficiency of egg production.

## ACKNOWLEDGMENTS

This research received a financial support from LIAN Development and Service.